\documentclass[acmsmall, screen]{acmart}

\usepackage{enumitem}
\usepackage{pifont}

\AtBeginDocument{%
  }

\setcopyright{acmcopyright}
\copyrightyear{2025}
\acmYear{2025}

\acmISBN{XXXXX-XXXX-XXXXX-XXXX-X}

\begin{document}

\title{Deep Learning Framework Testing via Heuristic Guidance Based on Multiple Model Measurements}

\author{Yinglong Zou}
\email{652023320004@smail.nju.edu.cn}
\orcid{0009-0006-9375-7417}
\affiliation{%
  \institution{State Key Laboratory for Novel Software Technology}
  \city{nanjing}
  \country{China}
}

\author{Juan Zhai}
\email{juanzhai@umass.edu}
\orcid{0000-0001-5017-8016}
\affiliation{%
  \institution{University of Massachusetts}
  \city{Amherst}
  \country{USA}
}

\author{Chunrong Fang}
\authornote{corresponding author}
\email{fangchunrong@nju.edu.cn}
\orcid{0000-0002-9930-7111}
\affiliation{%
  \institution{State Key Laboratory for Novel Software Technology}
  \city{Nanjing}
  \country{China}
}

\author{Yanzhou Mu}
\email{602022320006@smail.nju.edu.cn}
\orcid{0000-0003-1816-2246}
\affiliation{%
  \institution{State Key Laboratory for Novel Software Technology}
  \city{Nanjing}
  \country{China}
}

\author{Jiawei Liu}
\email{jw.liu@smail.nju.edu.cn}
\orcid{0000-0002-4930-9637}
\affiliation{%
  \institution{State Key Laboratory for Novel Software Technology}
  \city{Nanjing}
  \country{China}
}

\author{Zhenyu Chen}
\email{zychen@nju.edu.cn}
\orcid{0000-0002-9592-7022}
\affiliation{%
  \institution{State Key Laboratory for Novel Software Technology}
  \city{Nanjing}
  \country{China}
}

\renewcommand{\shortauthors}{Zou et al.}

\begin{abstract}
Deep learning frameworks serve as the foundation for developing and deploying deep learning applications. To enhance the quality of deep learning frameworks, researchers have proposed numerous testing methods using deep learning models as test inputs. However, existing methods predominantly measure model bug detection effectiveness as heuristic indicators, presenting three critical limitations. Firstly, existing methods fail to quantitatively measure model's operator combination variety, potentially missing critical operator combinations that could trigger framework bugs. Secondly, existing methods neglect measuring and heuristically guiding the model execution time, resulting in the omission of numerous models potential for detecting more framework bugs within limited testing time. Thirdly, existing methods overlook correlation between different model measurements, relying simply on single-indicator heuristic guidance without considering their trade-offs. To overcome these limitations, we propose DLMMM, the first deep learning framework testing method to include multiple model measurements into heuristic guidance and fuse these measurements to achieve their trade-offs. DLMMM firstly quantitatively measures model's bug detection performance, operator combination variety, and model execution time. After that, DLMMM fuses these measurements based on their correlation to achieve their trade-offs. To further enhance testing effectiveness, DLMMM designs multi-level heuristic guidance for test input model generation. We apply DLMMM to test three widely used deep learning frameworks (including TensorFlow, PyTorch, and MindSpore). The experimental results show that DLMMM outperforms state-of-the-art methods in effectiveness and efficiency. The test input models generated by DLMMM achieve a 250.13\% improvement in operator combination variety. The average model execution time of DLMMM is within 35\% of all state-of-the-art methods.
DLMMM detects 19 new crashes and 175 new NaN \& inconsistency bugs.
Among detected crashes, 16 have been confirmed by developers and two have been fixed in the following version.
\end{abstract}





\maketitle

\section{Introduction}
\label{introduction}
Deep Learning (DL) has been widely applied in many safety-critical tasks, such as autonomous driving \cite{autodrive, autodrive2, autodrive3}, medical diagnosis \cite{disease_diagnosis_1,disease_diagnosis_2}, and industrial safety monitoring \cite{industrialsafetymonitoring1,industrialsafetymonitoring2}. In recent years, researchers have shown increasing concerns about the security of DL applications \cite{DLSecurity1,DLSecurity2}. DL frameworks are the foundation for developing and deploying DL applications. Bugs in DL frameworks may potentially lead to application defects \cite{relation_DLapplication_framework}. Therefore, the quality of DL frameworks is crucial for the quality of DL applications. To ensure the quality of DL frameworks, researchers propose many framework testing methods. Among them, most generate DL models as test inputs \cite{lemon,muffin}. These methods execute generated models to invoke APIs provided by DL frameworks. To generate more effective models, these methods design measurements for each model's bug detection effectiveness and heuristically guide the following testing round to promote this measurement. These methods successfully detect many crashes and NaN \& inconsistency bugs in DL frameworks. However, there are still three critical limitations in the model measurement design. 

Firstly, the diversity of operator combinations within a test input model lacks the quantitative measurement. This diversity refers to the number of distinct substructures in a model, where each substructure represents a specific way of DL framework invocation. A greater variety of operator combinations implies more diverse ways in which the framework is invoked, ultimately leading to more comprehensive testing. Due to the complex structure of the model, quantitatively measuring the operator combinations variety is a significant challenge. Specifically, many DL models contain branches (e.g., ResNet \cite{resnet}), a special model substructure that enables multi-directional data flow within the model. Such models with branches cannot be simply represented as operator sequences, but require representation via directed acyclic graphs (DAGs). Accordingly, operator sequence-based variety measurements are no longer applicable to these models. To assess the variety of such structurally complex models, a new DAG-based measurement method needs to be designed.

Secondly, since the model execution time is not quantitatively measured or incorporated into heuristic guidance, the average model execution time is too long, which limits testing effectiveness. This limitation arises because a single model execution can trigger at most one DL framework bug—once triggered, this bug's impact propagates to all subsequent operators, preventing the exposure of other bugs during this execution. A longer average model execution time reduces the number of test input models that can be executed within the same testing time. As a result, some potentially bug-revealing models do not have the opportunity to be executed within the limited testing time, ultimately leading to some bugs not being successfully triggered. 

Thirdly, existing methods overlook the correlation between multiple model measurements, focusing narrowly on optimizing one single measurement without achieving overall optimization. For example, the bug detection effectiveness of a single model is a widely used measurement. It measures a model's effectiveness in triggering framework crashes and NaN \& inconsistency bugs. To maximize bug detection effectiveness, existing methods tend to excessively stack more operators into a model. However, this increases another important measurement, the average model execution time, thereby limiting the efficiency and effectiveness of the overall testing method.

To overcome these limitations, we quantitatively measure the model execution time and operator combination variety, and design measurement fusion based on the correlation between multiple model measurements. In this paper, we conduct an empirical study to investigate the effect of operator combination variety and model execution time, and the correlation between the above model measurements. We have three interesting findings:
\begin{itemize}[leftmargin = *]
    \item \textbf{Finding 1:} Higher operator combination variety contributes to increasing model's bug detection effectiveness.
    \item \textbf{Finding 2:} Higher operator combination variety contributes to increasing model's model execution time.
    \item \textbf{Finding 3:} Longer model execution time limits the efficiency of the overall testing method.
\end{itemize}

From Finding 1 and 3, higher operator combination variety and shorter model execution time are beneficial for DL framework testing. From Finding 2, a trade-off is required between increasing operator combination variety and reducing model execution time. Motivated by the above findings, we propose DLMMM, the first DL framework testing method that includes multiple model measurements into heuristic guidance. Firstly, DLMMM quantitatively measures the operator combination variety and model execution time, and adopts the widely used measurement for bug detection effectiveness. Secondly, DLMMM designs a measurement fusion based on the correlation between multiple model measurements to achieve their trade-offs. Thirdly, DLMMM designs a multi-level heuristic, mutation-based test input model generation based on the fused model measurements. In our model generation, the selection of the seed models and the model mutation rules are all heuristically guided.

We conduct experiments to evaluate the effectiveness and efficiency of DLMMM against four state-of-the-art methods, including LEMON \cite{lemon}, Muffin \cite{muffin}, COMET \cite{comet} and Gandalf \cite{gandalf}. The experimental results show that DLMMM is better than all state-of-the-art methods in effectiveness and efficiency. Specifically, the average operator combination variety increases by 250.13\%. The average model execution time of DLMMM is within 35\% of all baselines. DLMMM newly detects 19 crashes and 175 NaN \& inconsistency bugs. Among detected crashes, 16 are confirmed by developers, and two are fixed. 

To sum up, our main contributions are as follows.
\begin{itemize}
\item We conduct an empirical study to investigate the effect of operator combination variety and model execution time, and the correlation between above model measurements. As far as we know, we are first to quantitatively measure operator combination variety.
\item We are the first to propose the heuristic DL framework testing method based on multiple model measurements. This method designs a measurement fusion based on the correlation between multiple model measurements, as well as a multi-level heuristic, mutation-based test input model generation.
\item We implement our method as an open-sourced tool, named DLMMM, and conduct an evaluation on DLMMM to test three popular DL frameworks. The results show that our method newly detects 19 crashes and 175 NaN \& inconsistency bugs, which is the most among all baselines. In addition, DLMMM enhances operator combination variety by 250.13\% over the best baseline while maintaining an average model execution time within 35\% of state-of-the-art methods.

\end{itemize}

More information is available on Github: https://github.com/DeepLMMM/DLMMM.
\section{Background}

\subsection{Heuristic DL Framework Testing}
\label{heuristic dl framework testing}
DL framework testing aims to exercise the APIs (which is called operators) provided by DL frameworks. Many methods regard DL models and tensors as test inputs and conduct differential testing under different frameworks. Firstly, they design methods to generate test input models and tensors. Secondly, they execute generated test inputs on different DL frameworks. Thirdly, they detect bugs by comparing the output tensors from different DL frameworks. 

Heuristic guidance is an important component in DL framework testing, which is usually used to dynamically control the test input model generation. Many DL framework testing methods design heuristic guidance to enhance bug detection effectiveness. For example, COMET \cite{comet} applies MCMC \cite{MCMC} search algorithm to generate models covering more operator types. GraphFuzz \cite{fanggraphbased} proposes an operator-level coverage to increase the operator variety. Gandalf \cite{gandalf} applies context-free grammar and deep-Q network to generate test input models with higher model diversity. However, existing methods only focus on the variety of operator types, but fail to generate models with various operator combinations. In addition, the execution time of models generated by existing methods is usually too long. The above challenges motivate us to quantitatively measure operator combination variety and model execution time, and include these measurements into heuristic guidance.

Additionally, the heuristic guidance in existing methods only includes one model measurement, such as simply improving model's bug detection effectiveness. Existing methods fail to account for the trade-offs between multiple model measurements. The above challenges motivate us to fuse multiple measurements based on their relationship.

\subsection{Representation for Test Input Models in DL Framework Testing} 
\label{representation for models}
Proper model representation is a prerequisite for model generation. Because graphs clearly represent the relationship between operators and the data flow within a model, Directed Acyclic Graphs (DAG) are widely used to represent test input models in DL framework testing (e.g., LEMON \cite{lemon} and Muffin \cite{muffin}). Specifically, each DL model is represented as a DAG, corresponding to a tuple $<V_G, E_G>$. $V_G$ denotes the vertex set in the DAG, with each vertex in $V_G$ corresponding to a tensor. $E_G$ represents the edge set in the DAG, where each edge is a labeled directed edge corresponding to an operator in the model. Specifically, the label of each edge in $E_G$ represents the operator type, the starting vertex of the edge represents the input tensor of the operator, and the ending vertex represents the output tensor of the operator. Since the test input model contains only one input tensor and one output tensor, the DAG should include exactly one source vertex and one sink vertex. Furthermore, to ensure the validity of the model, the DAG should be a connected graph.

The branches in a model are well-represented in a DAG. Specifically, when a vertex in the DAG has an out-degree greater than one, it indicates the tensor corresponding to this vertex is used as the input for multiple operators. When a vertex in the DAG has an in-degree greater than one, it indicates that the outputs of multiple operators fuse into the tensor corresponding to this vertex.
\section{Empirical Study}
\label{empirical_study}

As mentioned in Section \ref{introduction}, researchers' subjective experience suggests that operator combination variety and model execution time are potentially important measurements that may influence bug detection performance. However, the impact of these measurements has not been quantitatively characterized. Moreover, existing studies have not explored the trade-offs between these measurements. To bridge this gap and lay the groundwork for designing DLMMM's heuristic guidance, we conduct an empirical study to investigate the effects of operator combination variety and model execution time, as well as the correlation between these model measurements.

\subsection{Empirical Study Setup}
\label{empirical_study_setup}

We execute four state-of-the-art DL framework testing methods, including LEMON \cite{lemon}, Muffin \cite{muffin}, COMET \cite{comet}, and Gandalf \cite{gandalf}. Firstly, we download their source code from the open-source repositories. Secondly, we run each method for six hours and save all generated models. No new bugs are detected in the last hour, indicating the convergence of the experiment. Thirdly, we execute saved models under the same environment to test DL frameworks, which include TensorFlow, PyTorch, and MindSpore. For fairness, all test input tensors are randomly generated and kept the same. During the model execution, we record each model's bug detection effectiveness, operator combination variety, and model execution time (see Section \ref{the measurement for models} for their calculation).

Based on recorded model measurements, we investigate their correlation. Considering the potential non-normal distribution of the data and the presence of outliers, we use the Spearman correlation coefficient to describe their relationship. The Spearman correlation coefficient is a measurement used for evaluating the correlation between two variables, with a range of [-1, 1] \cite{Spearman_correlation_coefficient}. The calculation for the Spearman correlation coefficient \(\rho\) is:
$$\rho = 1 - \frac{6 \sum_{i=1}^n d_i^2}{n(n^2 - 1)}$$

where \(n\) represents the number of data points, and \(d_i\) represents the difference between the ranks of corresponding values of the two variables. A positive Spearman correlation coefficient indicates a positive monotonic correlation between the two variables, while a negative coefficient indicates a negative monotonic correlation. The larger the absolute value of the Spearman correlation coefficient, the stronger the monotonic correlation between the two measurements. Generally, when the absolute value of the Spearman correlation coefficient is greater than 0.2, a relatively strong monotonic correlation between the two variables is considered to exist \cite{Spearman_correlation_coefficient}.


\subsection{Model Measurements}
\label{the measurement for models}

In this section, we introduce three model measurements adopted in our empirical study, including bug detection effectiveness (Section \ref{bug detection effectiveness measurement}), operator combination variety (Section \ref{operator combination variety measurement}), and model execution time (Section \ref{model execution time measurement}).

\subsubsection{Bug Detection Effectiveness Measurement}
\label{bug detection effectiveness measurement}
Bug detection effectiveness is a widely used measurement by many DL framework testing methods (e.g, Muffin \cite{muffin}, Gandalf \cite{gandalf}). Following these methods, we continue to use this measurement to evaluate the framework testing effectiveness of a single model. The calculation for bug detection effectiveness differs depending on whether the model triggers a crash or NaN bug. Specifically, when the model successfully triggers a crash or NaN bug, the bug detection effectiveness (denoted as $performance$) is the average value of all the data in the test input tensor. If the model fails to trigger a crash or NaN bug, the bug detection effectiveness takes the value of $inconsistency$. $Inconsistency$ is proposed based on the idea of differential testing, which is calculated based on the difference of output tensors under multiple different frameworks. The $inconsistency$ between DL framework $i$ and DL framework $j$ (denoted as $inconsistency_{ij}$) is recorded and calculated as follows.
\ding{192} Provide the same input tensor, perform inference under DL framework $i$ and $j$, and record their result tensor as $result_i$ and $result_j$. \ding{193} Calculate the difference between $result_i$ and $result_j$ as $result\_diff_{ij}$. \ding{194} Calculate the maximum value among all numbers in $result\_diff_{ij}$ as $inconsistency_{ij}$.

\subsubsection{Operator Combination Variety Measurement}
\label{operator combination variety measurement}
Operator-level variety (including operator variety and operator combination variety) has attracted widespread attention from DL framework testing researchers. Many researchers (e.g., Gandalf \cite{gandalf}, Muffin \cite{muffin}) qualitatively believe that since many DL framework bugs are closely related to specific model substructures, the higher the variety of operators and operator combinations in the model, the more effective the framework testing will be. However, although operator combination variety is crucial for enhancing the effectiveness of framework testing, designing a universal quantitative measurement for operator combination variety is a highly challenging task. The main challenge lies in the complexity of model structures, as some models include branches. Existing methods have made some attempts but have failed to overcome this challenge. For example, COMET \cite{comet} measures operator combination variety based on the operator sequence of the model. Operator sequence can only describe single-directional data flow within the model, instead of representing model with branches (multi-directional data flow). A more universal quantitative measurement is urgently needed. 

To overcome this challenge, we propose the measurement $variety\_degree$ for operator combination variety. $Variety\_degree$ is calculated based on the number of subgraphs contained in the DAG of the model. 
The detailed definition and calculation for $variety\_degree$ is as follows:
Firstly, set the number of operators allowed in a subgraph, called $depth$. Secondly, find all subgraphs, called $motifs$, in the DAG representation of the model, where the number of operators in each subgraph equals to $depth$. Thirdly, remove all isomorphic subgraphs from $motifs$, which are considered as redundant subgraphs. Finally, count the remaining subgraphs in $motifs$, called $variety\_degree$.

$variety\_degree$ is a universal measurement that reflects the variety of both single operators and operator combinations. Specifically, when $depth = 1$, $variety\_degree$ describes the variety of single operators in the model; And when $depth > 1$, $variety\_degree$ describes the variety of operator combinations in the model. Since operator combinations with fewer operators often have higher practical reproducible value, and it is too time-consuming to set the $depth$ larger than 2, the $depth$ is set to 2.

\subsubsection{Model Execution Time Measurement}
\label{model execution time measurement}
The execution time of a model, denoted as $time$, is a measurement recorded during test runs. The model execution time is the primary time overhead for each round of framework testing, reflecting the efficiency of the framework testing. Generally, the longer the execution time of a single model, the longer the time each round of testing takes, and the fewer test input models are executed. Some models that may trigger DL framework bugs may lose the opportunity to be executed, thereby reducing the effectiveness and efficiency of the method. Due to the lack of awareness of the correlation between the bug detection effectiveness of a single model and the model execution time, the execution time is overlooked in existing DL framework testing methods.

\subsection{Research Questions}

In this study, we investigate three research questions:
\begin{itemize}[leftmargin = *]
 \item \textbf{RQ1:} How does operator combination variety affect DL framework testing?
  \item  \textbf{RQ2:} Is there any correlation between operator combination variety and model execution time?
  \item \textbf{RQ3:} How does model execution time affect DL framework testing?
\end{itemize}

In RQ1 and RQ3, we discuss the effect of operator combination variety and model execution time. In RQ2, we discuss the correlation between the above two model measurements to illustrate the necessity of our measurement fusion.

\subsection{Empirical Results for RQ1 \& RQ2}

The Spearman correlation coefficient among model measurements is shown in Table \ref{empirical_study_relationship_table}. The first column represents the method names. The second column shows the Spearman correlation coefficient between the operator combination variety and bug detection effectiveness. The third column shows the Spearman correlation coefficient between the operator combination variety and model execution time. It is worth noting that LEMON generates only four unique models in six hours, which is too small to calculate Spearman correlation coefficient. Therefore, we mark the empirical results of LEMON as NaN.

\begin{table}[htpb]
  \caption{Spearman Correlation Coefficient among Model Measurements}
  \label{empirical_study_relationship_table}
  \centering
  \begin{tabular}{ccc}
  \hline
   & \textbf{Variety \& Performance} & \textbf{Variety \& Time}\\
  \hline
    LEMON  & NaN & NaN\\
    Muffin  & 0.43 & 0.56\\
    COMET  & 0.44 & 0.38\\
    Gandalf  & 0.52 & 0.79\\
    \hline
  \end{tabular}
\end{table}


From the above results, we have two interesting findings:

\textbf{Finding 1 (Answer to RQ1): Higher operator combination variety contributes to increasing bug detection effectiveness of a model.} In the second column of Table \ref{empirical_study_relationship_table}, the Spearman correlation coefficient between operator combination variety and bug detection effectiveness exceeds 0.2 in all methods (0.43 in Muffin, 0.44 in COMET, and 0.52 in Gandalf), which indicates a positive correlation between the above two model measurements. In conclusion, higher operator combination variety contributes to improving the bug detection effectiveness.

\textbf{Finding 2 (Answer to RQ2): Higher operator combination variety contributes to increasing the model execution time of a model.} In the last column of Table \ref{empirical_study_relationship_table}, the Spearman correlation coefficient between model execution time and operator combination variety exceeds 0.2 in all methods (0.56 in Muffin, 0.38 in COMET, and 0.79 in Gandalf), which indicates a positive correlation between model execution time and operator combination variety. This correlation validates our intuition: the higher the operator combination variety, the more complex the computation performed by the model, and the longer the model execution time.

\subsection{Empirical Results for RQ3}
To investigate the effect of model execution time on DL framework testing, we sort the models generated by each baseline in descending order of model execution time and divide these models into two groups. One group consists of models with execution times in the higher 50\% of all models generated by a baseline, named ``larger model". The other group consists of models with execution times in the lower 50\%, named ``smaller model". We record the total execution time for each group, the average execution time per model in each group, the total number of bugs detected by each group, and the average time to detect each bug. The results for COMET and Gandalf are shown in Table \ref{empirical_study_time_comet_table} and Table \ref{empirical_study_time_gandalf_table}, respectively. Since LEMON only detects one DL framework bug and Muffin does not detect any DL framework bugs, we exclude them in RQ3.

\begin{table*}[htpb]
  \caption{Effect of Model Execution Time in COMET}
  \label{empirical_study_time_comet_table}
  \centering
  \begin{tabular}{ccccc}
  \hline
   & \textbf{Total Time (s)} & \textbf{Time Per Model (s)} & \textbf{Bug Number} & \textbf{Time Per Bug (s)} \\
  \hline
    smaller model  & 6955.05 & 92.73 & 7 & 993.58\\
    larger model  &  14644.95 & 195.27 & 12 & 1195.41\\
    \hline
  \end{tabular}
\end{table*}

\begin{table*}[htpb]
  \caption{Effect of Model Execution Time in Gandalf}
  \label{empirical_study_time_gandalf_table}
  \centering
  \begin{tabular}{ccccc}
  \hline
   & \textbf{Total Time (s)} & \textbf{Time Per Model (s)} & \textbf{Bug Number} & \textbf{Time Per Bug (s)} \\
  \hline
    smaller model  & 2538.85 & 8.70 & 23 & 110.38\\
    larger model  &  19061.15 & 65.06 & 24 & 794.21\\
    \hline
  \end{tabular}
\end{table*}

From the above results, we have the following findings:

\textbf{Finding 3 (Answer to RQ3): Longer model execution time limits the efficiency of DL framework testing.} In table \ref{empirical_study_time_comet_table} and \ref{empirical_study_time_gandalf_table}, although the total bug number triggered by larger models are more than that triggered by smaller models (12 greater than 7 in COMET, 24 greater than 23 in Gandalf), the bug detection efficiency of smaller models exceeds that of larger models. Specifically, in COMET, smaller models take an average of 993.58 seconds to detect a bug, less than the 1195.41 seconds taken by larger models. In Gandalf, smaller models take an average of 110.38 seconds to detect a bug, less than the 794.21 seconds taken by larger models. The advantage of smaller models in bug detection efficiency comes from their shorter execution time (92.73 seconds shorter than 195.27 seconds in COMET, 8.70 seconds shorter than 65.06 seconds in Gandalf). In conclusion, longer model execution time limits the bug detection efficiency.
\section{METHODOLOGY}

\subsection{Overview}
\label{overview}
Our findings inspire us to generate test input models with higher operator combination variety and lower model execution time, and make trade-offs between these model measurements. In this work, we propose DLMMM, a DL framework testing method via heuristic guidance based on multiple model measurements, including bug detection effectiveness, operator combination variety, and model execution time. Based on the correlation between these measurements, DLMMM designs measurement fusion to calculate a heuristic indicator from these measurements and designs multi-level heuristic test input model generation to further enhance testing effectiveness.

The overall workflow of DLMMM is shown in Figure \ref{Workflow of DLMMM}. Firstly, DLMMM generates test input tensors based on the pre-specified tensor shape (Section \ref{input tensor generation}). Secondly, DLMMM generates a trivial model based on the model configuration and randomly mutates the trivial model to construct seed models (Section \ref{seed model generation}). Thirdly, DLMMM selects and mutates generated seed models to construct test input models (Section \ref{input model generation}). Fourthly, DLMMM adopts differential testing under different frameworks, using generated tensors and models as test inputs (Section \ref{differential testing}). DLMMM detects crashes and NaN \& inconsistency bugs by analyzing the result of differential testing. In differential testing, DLMMM records three measurements for each test input model, including bug detection effectiveness, operator combination variety, and model execution time. At Last, based on recorded measurements, DLMMM designs a heuristic guidance, which fuses model measurements and provides the feedback to test input model generation of next round (Section \ref{feedback}).

\begin{figure*}[htpb]
    \centering
    \includegraphics[width=0.95\textwidth]{Figure/overview.pdf}
    \caption{Workflow of DLMMM}
    \label{Workflow of DLMMM}
\end{figure*}

\subsection{Test Input Tensor Generation}
\label{input tensor generation}
Tensor is an important component of test inputs in DL framework testing. Following existing methods  (e.g., LEMON \cite{lemon}), DLMMM designs two test input tensor generation modes. In one mode, DLMMM loads tensors from publicly available datasets. DLMMM supports all datasets adopted by existing methods, which include MNIST, Fashion-MNIST, CIFAR-10, ImageNet, Sine-Wave, and Stock-Price. In another mode, DLMMM randomly generates test input tensors based on the tensor shape. The tensor shape is specified before testing begins, which is in the format of $NCHW$  ($batch$, $channel$, $height$, $width$).

\subsection{Seed Model Generation}
\label{seed model generation}
As the foundation of model generation, diverse seed models contribute to generating diverse test input models. Existing methods, however, predominantly apply publicly available models or model templates as seeds, limiting the seeds' diversity. To facilitate generating diverse test input models, DLMMM proposes a mutation-based seed model generation method, which generates various seed models without relying on publicly available models or templates.

\subsubsection{Model Mutation}
\label{mutation}
Model mutation is used to generate new DL models. Existing methods design mutation rules (e.g., insert operator, remove operator) and heuristically select which mutation rule to apply. These methods fail to heuristically control the specific implementation of each mutation rule. To overcome this limitation, DLMMM designs a double-level heuristic mutation process based on the DAG representation of the model (introduced in Section \ref{representation for models}). In DLMMM's mutation process, mutation rules are implemented as replacing edges in the DAG (e.g., replacing an edge representing the operator $None$ with the edge representing another operator corresponds to the mutation rule ``insert operator" in existing methods). By the following heuristic mutation process, in DLMMM, the heuristic indicator not only controls the probability of selecting each mutation rule, but also controls the specific implementation of each mutation (e.g., the probability of inserting or removing each type of operator). The detailed mutation process is as follows. 
\ding{192} Randomly sample a predecessor vertex $i$  ($i$ is not the sink). 
\ding{193} Randomly sample a successor vertex $j$  ($j > i$).
\ding{194} Sample an operator $o$ based on the $weight$ of each operator. The calculation of $weight$ is introduced in Section \ref{feedback}. The probability $p$ of each operator being sampled is: 
    $$p = \frac{weight}{\sum_{k=1}^{n}weight_k}$$
where $n$ is the number of operators under test.
\ding{195} Replace the operator from the vertex $i$ to the vertex $j$ with the operator $o$.
\ding{196} Check whether the newly generated model is valid (see Section \ref{representation for models} for requirements for a valid model). If not, return to step 1.

\subsubsection{Mutation-Based Seed Model Generation}
\label{mutation-based seed model generation} DLMMM generates test input models from sketch. Specifically, before the method begins, the pre-specified model configuration determines the number of operators contained in each model. After that, DLMMM firstly initializes an empty seed model pool, and generates an seed model without any operators, which is called the trivial model. The DAG representation for this trivial model is an operator chain from the source vertex to the sink vertex composed of the operator $identity$. Then, DLMMM randomly selects a seed model from the seed model pool, mutate it according to the mutation described in Section \ref{mutation}, and add the newly generated model into the seed model pool. Finally, DLMMM repeats the above model selection and mutation until the seed model pool reaches a pre-specified size (configured before testing begins).

\subsection{Test Input Model Generation}
\label{input model generation}
DLMMM generates new test input models by mutating seed models in the seed model pool (constructed in Section \ref{seed model generation}). In DLMMM, seed model selection and model mutation are both heuristically guided by the heuristic indicator $fitness$, which is calculated based on the recorded model measurements (will be introduced in Section \ref{feedback}). In seed model selection, DLMMM uses a tournament algorithm \cite{tournament} to select $k$ seed models from the seed model pool based on their $fitness$. $K$ is the specified before DLMMM begins, which is usually set to 1 to promote algorithm's efficiency. The tournament algorithm is performed as follows. \ding{192} Initialize an empty winner seed model pool. \ding{193} Randomly sample $k$ seed models from the seed model pool. \ding{194} Select the seed model with the highest $fitness$ from the subset selected in Step \ding{193}. \ding{195} Add the selected seed model to the winner seed model pool and remove it from the seed model pool to avoid reselection. \ding{196} Repeat Step \ding{193} - \ding{195} until $k$ models are added into the winner seed model pool. The winner seed model pool is the output of the tournament algorithm.

Based on selected seed models, DLMMM applies a double-level heuristic mutation on the selected seed models to generate new test input models (the mutation is introduced in Section \ref{mutation}). These newly generated models are then used as test inputs to exercise different DL frameworks.

\subsection{Differential Testing}
\label{differential testing}
DLMMM adopts differential testing among different DL frameworks. The test inputs include the test input tensors generated in Section \ref{input tensor generation} and the test input models generated in Section \ref{input model generation}. DLMMM executes the above test inputs under different DL frameworks and records execution logs and output tensors. By analyzing logs and output tensors, DLMMM targets both non-numerical bugs and numerical bugs. Specifically, the non-numerical bugs include crashes. The numerical bugs include Not-A-Number (NaN) bugs, and inconsistency bugs.

\textbf{Crashes.} DLMMM keeps logs during the execution to detect crashes. Two researchers analyze exception messages recorded in these logs, respectively, to confirm the root cause of crashes and avoid redundancy.

\textbf{NaN \& inconsistency bugs.} DLMMM designs a voting mechanism to figure out which framework triggers NaN \& inconsistency bugs. Specifically, NaN bugs are detected when the output tensor of a framework contains NaN while the output tensors of other frameworks do not. Inconsistency bugs are detected when two $inconsistency$ (see Section \ref{bug detection effectiveness measurement}) calculated based on the output tensor under the same framework both exceed the threshold $\epsilon$. When more than two $inconsistency$ exceed $\epsilon$, DLMMM attributes this inconsistency bug to the common framework related to the largest two $inconsistency$. To ensure fairness in the experiment and better eliminate errors caused by computational randomness, DLMMM sets the value of the threshold to the maximum value among all existing methods (e.g., Predoo \cite{predoo}, Gandalf \cite{gandalf}, and Muffin \cite{muffin}). The value of the threshold $\epsilon$ is 0.15. In addition, an NaN \& inconsistency bug may be triggered by multiple the same models, which leads to redundancy. To avoid redundancy, DLMMM analyzes the model structures and removes redundant models.

In the differential testing, DLMMM calculates and records three measurements for the newly generated model, including bug detection effectiveness, model execution time, and operator combination variety (see Section \ref{the measurement for models} for their calculation method). In the next section, we will introduce how DLMMM guides the model generation based on these measurements.

\subsection{Heuristic Guidance}
\label{feedback}
\label{measurement fusion}
\label{heuristic guidance to following rounds}
Heuristic guidance is an important component in DL framework testing, which is usually used to control test input model generation. As found in Section \ref{empirical_study}, there are correlations between multiple model measurements. To achieve a trade-off between multiple model measurements and improve the overall testing effectiveness, DLMMM designs a measurement fusion based on the correlations between model measurements. Specifically, DLMMM fuse bug detection effectiveness, operator combination variety, and model execution time into the heuristic indicator $fitness$, which is expected to meet the following five requirements.
\ding{192} $Fitness$ should reflect the overall level of the models in bug detection effectiveness, model execution time and operator combination variety. \ding{193} $Fitness$ should exhibit positive correlation with bug detection effectiveness and operator combination variety, and should exhibit negative correlation with model execution time. \ding{194} $Fitness$ should capture the contrast intensity in each measurement among different models. \ding{195} $Fitness$ should capture the correlation between different measurements \ding{196} $Fitness$ should eliminate dimensional differences among different measurements.

To meet the above five requirements, DLMMM's measurement fusion is designed based on CRITIC \cite{CRITIC}. CRITIC is a classical objective weighting method, which is used to determine the objective weights of all measurements based on their contrast intensity and conflict. Specifically, contrast intensity is calculated by the standard deviation, and conflict between measurements is based on their correlation. In DLMMM, the $fitness$ of a newly generated model is calculated based on the model measurements of itself and all models generated before, which is recorded in differential testing (see Section \ref{differential testing}). For convenience in calculation, we place all recorded model measurements into a matrix named $judge\_matrix$. The rows in the $judge\_matrix$ represent the measurement values of each model. The first column in the $judge\_matrix$ represents bug detection effectiveness, denoted as $performance\_ column$. The second column represents operator combination variety, denoted as $variety\_column$. The third column represents model execution time, denoted as $time\_column$. The detailed calculation for $fitness$ is as follows:

\ding{192} Replace each value in $time\_column$ by its reciprocal to maintain a negative correlation between model execution time and the fused heuristic indicator.

\ding{193} Normalize $performance\_column$, $variety\_column$, and $time\_column$ as follows:
$${z}_{pq}=\frac {{x}_{pq}} {\sqrt {\sum ^{n}_{p=1} {{x}^{2}_{pq}}}}$$
In this formula, $p$ represents the row index in each column. $q$ represents the column, which is in the set $\{performance\_column, variety\_column, time\_column\}$. ${z}_{pq}$ represents the normalized value, while ${x}_{pq}$ represents the original, non-normalized value.

\ding{194} Calculate contrast intensity of each column ${\sigma }_{q}$ as follows:
$$ {\sigma }_{q}=\sqrt {\frac {\sum ^{m}_{p=1} { ({z}_{pq}-\bar{z})}} {m-1}} $$
In this formula, $p$ and $q$ represent the row and column index, respectively. ${z}_{pq}$ represents the value in each column. $m$ represents the number of values contained in each column, while $\bar z$ represents the average value in each column.

\ding{195} Calculate the conflict between columns $f_q$ as follows:
$$ {f}_{q}=\sum { (1-{r}_{q})} $$
In this formula, the meanings of $q$ are the same as in Step 2. $r_{q}$ is the Spearman correlation coefficient, which reflects the correlation between measurements.

\ding{196} Based on the contrast intensity ${\sigma }_{q}$ and conflict $f_q$ calculated in Step 3 and 4, compute the information carrying capacity $c_q$ of each column as follows:
$$C_q=\sigma_q f_q$$

\ding{197} Calculate the weight for each column $w_q$ based on the information carrying capacity $c_q$ calculated in Step 5 as follows:
$$ {w}_{q}=\frac {{c}_{q}} {\sum ^{n}_{q=1} {{c}_{q}}} $$
In this formula, $n$ represents the number of measurements.

\ding{198} Calculate the heuristic indicator $fitness$ of the newly generated model based on its original data $x_{q}$ in the $judge\_matrix$ and the weights of each column $w_q$ (calculated in Step 6) as follows:
$$ fitness=\sum ^{n}_{q=1} {{w}_{q}}{x}_{q} $$.

The $fitness$ calculated using the above measurement fusion depicts the contrast intensity of the data from different models within each model measurement, and the correlation between multiple measurements. In addition, the above measurement fusion eliminates the dimensional differences between different model measurements. Therefore, $fitness$ reflects the overall performance of generated models in bug detection effectiveness, operator combination variety, and model execution time. Based on the $fitness$ of the seed model and the newly generated model, DLMMM calculates the contribution of this mutation $\Delta fitness$ as follows:
$$\Delta fitness = fitness_{new} - fitness_{seed}$$
where $fitness_{new}$ is the $fitness$ of the newly generated model and $fitness_{seed}$ is the $fitness$ of the seed model. Based on $\Delta fitness$, DLMMM updates the $weight$ of each operator, which is used to calculate the probability being sampled in mutation. The update formula is as follows:
    $$weight_1 = weight_0 + \Delta fitness$$
where $weight_1$ and $weight_0$ is the $weight$ after and before updating, respectively.
\section{Evaluation}
\label{evaluation}

To evaluate the effectiveness of DLMMM in detecting DL framework bugs, we compare DLMMM with four baselines: LEMON \cite{lemon}, Muffin \cite{muffin}, COMET \cite{comet}, and Gandalf \cite{gandalf}. Because there are no publicly available codes for graphFuzz \cite{fanggraphbased}, we do not select it as our baseline. All codes and experimental results are available in our public repository.

\subsection{Experimental Setup}
\label{experimental setup}

We test three popular DL frameworks: TensorFlow \cite{tensorflow}, PyTorch \cite{pytorch}, and MindSpore \cite{mindspore}. The default software versions are: TensorFlow v2.9.0, PyTorch v1.12.0, and MindSpore v2.1.0. The workstation is equipped with the Ubuntu 22.04 operating system and NVIDIA RTX 4090D GPU, and an Intel Core Processor (Skylake) (16cores, 2.0GHz). The experimental time on each dataset (see Section \ref{input tensor generation}) is six hours. No new bugs are detected during the last hour of each experiment, demonstrating the convergence. The threshold $\epsilon$ for detecting inconsistency bugs is 0.15 (see Section \ref{differential testing}).

\subsection{Research Questions}
\label{research questions}

In this paper, we perform the evaluation along with the following research questions.

\begin{itemize}
\item \textbf{RQ4:} How does DLMMM perform in detecting DL framework bugs?
\item  \textbf{RQ5:} Can DLMMM generate test input models effectively?
\item \textbf{RQ6:} To what extent do the model measurements and their fusion in heuristic guidance contribute to the bug detection effectiveness of DLMMM?
\end{itemize}

In RQ4, we evaluate the effectiveness of DLMMM in detecting crashes and NaN \& inconsistency bugs. In RQ5, we evaluate the test input models generated by DLMMM. In RQ6, we conduct an ablation study to analyze the contribution of novel designs in heuristic guidance.

\subsection{RQ4: Effectiveness of Bug Detection}
\label{RQ4_Effectiveness_Bug}

We evaluate DLMMM's bug detection effectiveness from two aspects,  effectiveness in detecting crashes and NaN \& inconsistency bugs, respectively.

\subsubsection{Crashes}
\begin{figure*}[htpb]
    \centering
    \includegraphics[width=0.9\textwidth]{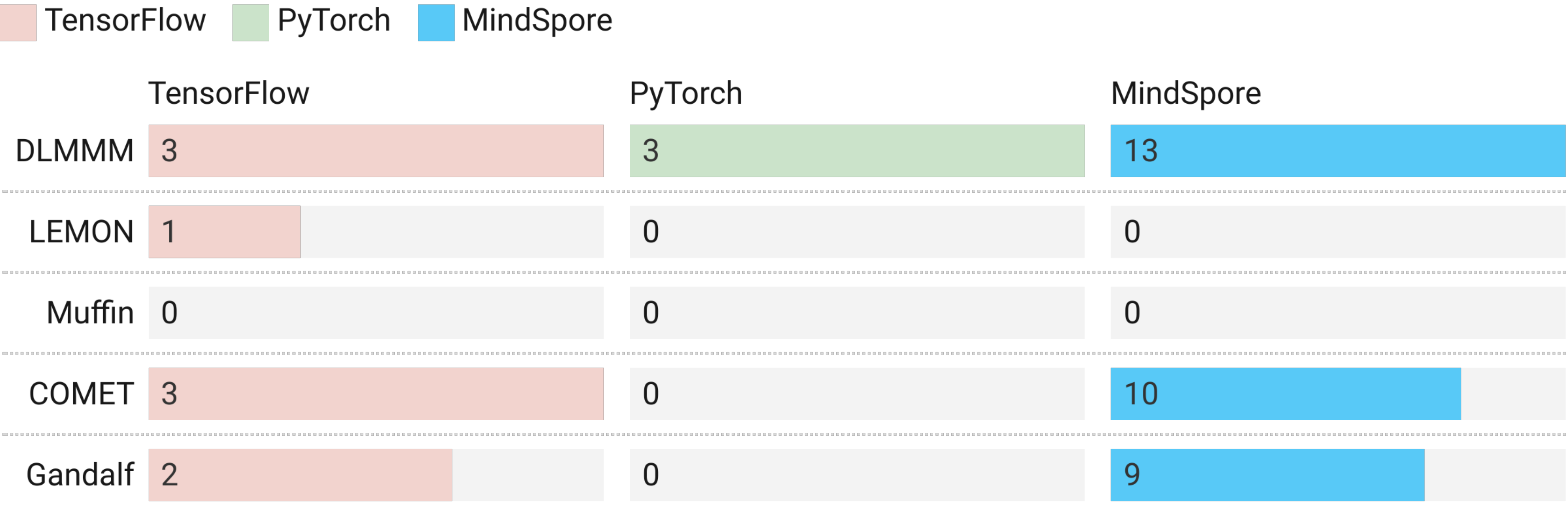}
    \caption{Number of Unique Crashes Detected by Each Method}
    \label{chart_crash_detection}
\end{figure*}

As introduced in Section \ref{differential testing}, DLMMM records the logs while executing test input models to detect crashes. Since one crash may be repeatedly triggered on different datasets, two researchers analyze the logs respectively to remove redundant crashes. Table \ref{table_crash_detection} shows the crashes detected by DLMMM under each dataset. In this table, the first column represents the adopted dataset, and the following three columns represent the number of crashes detected under TensorFlow, PyTorch, and MindSpore. Figure \ref{chart_crash_detection} shows the total number of crashes triggered on all datasets after removing redundancy. In this figure, the first column represents the method name, and the following three columns represent the number of crashes detected under each tested DL framework. The experimental results show that DLMMM newly detects 19 crashes, including 3 in TensorFlow, 3 in PyTorch and 13 in MindSpore, which is the most among the four baselines. Specifically, only one crash is triggered by LEMON; 13 are by COMET, and 11 are by Gandalf. Muffin does not trigger any crashes. We have reported all the detected crashes to the public community. Among them, 16 of 19 have been confirmed by developers and two of 19 are fixed in the updated version. 

In addition, we analyze the root causes behind detected crashes. In TensorFlow, one crash arises from failure in initialization. Another two crashes are due to unsupported implementation (e.g., unsupported operator $sparse\_mat\_mul$ and unsupported parameter $dilations$ for the operator $Convolution$). In PyTorch, one crash arises from the failure in creating a descriptor. Another two crashes are due to unsupported implementation. In MindSpore, one crash arises from unsupported implementation in the operator $PRELU$ in the CPU environment. One crash arises from the failure in creating a descriptor. Another 11 crashes are caused by the failure in memory allocation. This issue leads to crashes during executing multiple operators, such as $Convolution$ and $Depthwise\_Convolution$.

\begin{table}[htpb]
  \caption{Number of Unique Crashes Detected by DLMMM } 
  \label{table_crash_detection}
  \centering
  \begin{tabular}{cccc}
  \hline
  &\textbf{TensorFlow} & \textbf{PyTorch} & \textbf{MindSpore}\\  \hline
    Random & 3 & 1 & 9\\
    MNIST & 2 & 0 & 4\\
    Fashion-MNIST & 2 & 0 & 0\\
    CIFAR-10 & 0 & 0 & 3\\
    ImageNet & 2 & 2 & 10\\
    Sine-Wave & 0 & 0 & 2\\
    Stock-Price & 0 & 0 & 2\\
    \hline
  \end{tabular}
\end{table}

\subsubsection{NaN \& Inconsistency Bugs}

\begin{figure*}[htpb]
    \centering
    \includegraphics[width=0.9\textwidth]{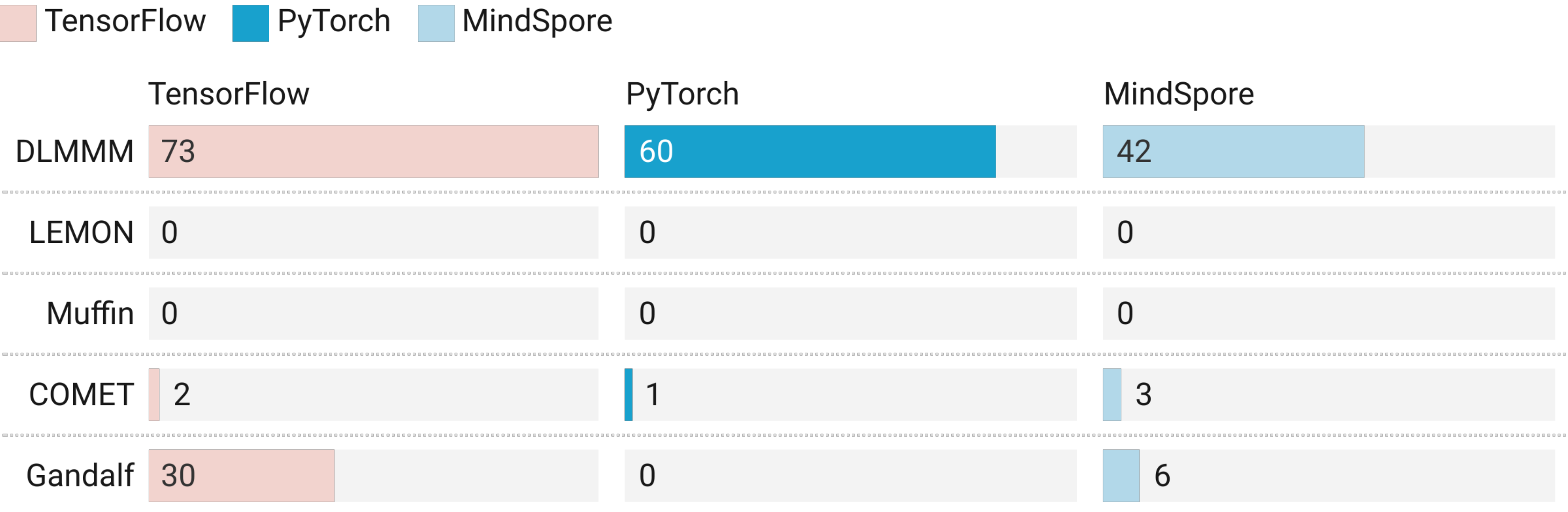}
    \caption{Number of Unique Detected NaN \& Inconsistency Bugs}
    \label{chart_inconsistency_detection}
\end{figure*}

Table \ref{table_inconsistency_detection} shows the number of NaN \& Inconsistency bugs triggered on all datasets. In this table, the first column represents the adopted dataset, and the following three columns represent the number of NaN \& Inconsistency bugs detected under each tested DL framework. This table shows NaN \& inconsistency bug is successfully triggered in almost all datasets (except MNIST). The small test input tensor size in MNIST (only 48 in height and width dimension) limits the bug detection effectiveness. Figure \ref{chart_inconsistency_detection} shows the number of unique NaN \& inconsistency bugs detected by DLMMM and all baselines. In this figure, the first column represents the method name, the following three columns represent the same meaning as Table \ref{table_inconsistency_detection}. The experimental results show that DLMMM triggers most NaN \& inconsistency bugs compared to four baselines. Specifically, DLMMM triggers 175 NaN \& inconsistency bugs in total, including 73 under TensorFlow, 60 under PyTorch, and 42 under MindSpore. LEMON and Muffin do not detect any NaN \& inconsistency bugs at all. COMET and Gandalf detect six and 36 NaN \& inconsistency bugs, respectively, under all these frameworks. In conclusion, DLMMM performs best in detecting NaN \& inconsistency bugs compared to all baselines under all frameworks.

In addition, we analyze the root causes behind detected NaN \& inconsistency bugs. We find that these bugs mainly arise from three root causes. Firstly, error arises during the calculation of each operator and accumulates during model inference, leading to NaN \& inconsistency bugs. Secondly, the fluctuation in numerical value caused by randomness usually leads to NaN \& inconsistency bugs. For example, many NaN \& inconsistency bugs are triggered by the operator $Dropout$, which brings randomness to the calculation. Thirdly, the implementation difference between different frameworks usually leads to NaN \& inconsistency bugs. For example, the padding modes for the operator $BatchNorm$ are different in PyTorch and MindSpore, which leads to an inconsistency bug when calling $torch.nn.Batchnorm2d ()$ and $mindspore.nn.Batchnorm2d ()$.

\begin{table}[htpb]
  \caption{Number of Unique NaN \& Inconsistency Bugs Detected by DLMMM}
  \label{table_inconsistency_detection}
  \centering
  \begin{tabular}{cccc}
  \hline
  &
  \textbf{TensorFlow} & \textbf{PyTorch} & \textbf{MindSpore}\\
  \hline
    Random & 7 & 26 & 6\\
    MNIST & 0 & 0 & 1\\
    Fashion-MNIST & 34 & 3 & 2\\
    CIFAR-10 & 6 & 27 & 5\\
    ImageNet & 1 & 1 & 1\\
    Sine-Wave & 23 & 2 & 6\\
    Stock-Price & 2 & 1 & 21\\
    \hline
  \end{tabular}
\end{table}

\begin{center}
\fcolorbox{black}{lightgray}{\parbox{.95\linewidth}{Answer to RQ4: DLMMM has an excellent performance in detecting crashes and NaN \& inconsistency bugs. DLMMM detects 19 crashes and 175 NaN \& inconsistency bugs, which is the most among all four baselines.}}
\end{center}

\subsection{RQ5: Effectiveness of Test Input Model Generation}
\label{RQ5_Effectiveness_Model}

To evaluate the effectiveness of test input model generation, we investigate the operator combination variety and model execution time of generated test input models. 

\subsubsection{Operator Combination Variety}
\label{RQ5_operator_combination_variety}
\begin{figure*}[htpb]
    \centering
    \includegraphics[width=0.9\textwidth]{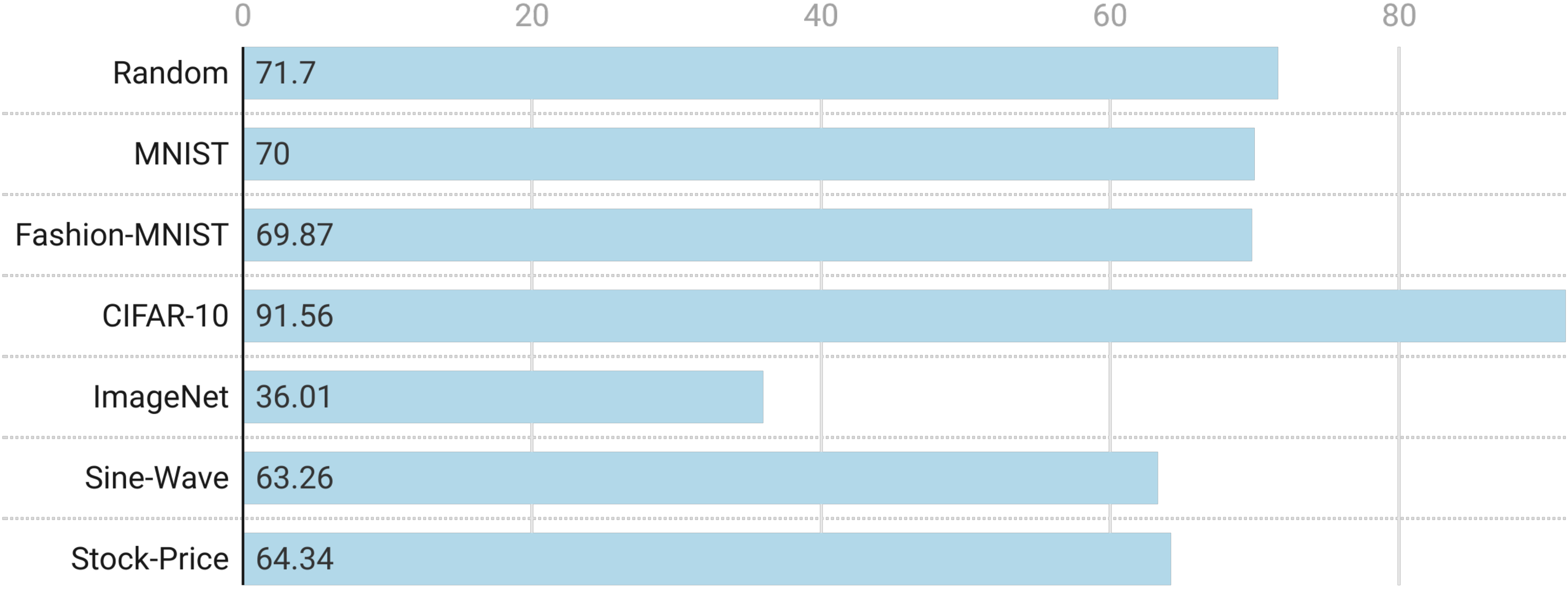}
    \caption{Operator Combination Variety of DLMMM}
    \label{chart_operator_combination_variety_dataset}
\end{figure*}

\begin{figure*}[htpb]
    \centering
    \includegraphics[width=0.9\textwidth]{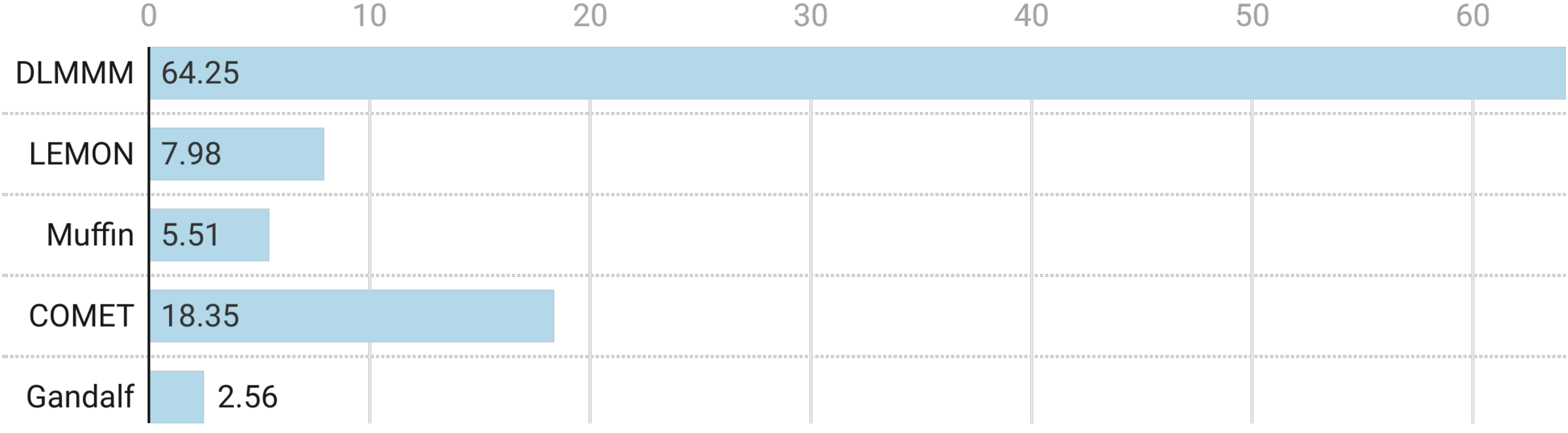}
    \caption{Comparison between DLMMM and Baselines in Operator Combination Variety}
    \label{chart_operator_combination_variety_method}
\end{figure*}

Figure \ref{chart_operator_combination_variety_dataset} shows the average operator combination variety of DLMMM on each dataset. In this figure, the text on the axis represents the datasets, and the numbers on the bars represent the average operator combination variety of each dataset (measured by $variety\_degree$ introduced in Section \ref{operator combination variety measurement}). Figure \ref{chart_operator_combination_variety_method} shows the comparison between DLMMM and baselines in average operator combination variety in all datasets. In this figure, the text on the axis represents the method name, and the numbers on the bars represent the average operator combination variety of each method. The experimental results show that the average operator combination variety of test input models generated by DLMMM is 64.25, which greatly exceeds 7.98 in LEMON, 5.51 in Muffin, 18.35 in COMET, and 2.56 in Gandalf. Compared to the best performance of all baselines, DLMMM achieves a 250.13\% improvement in operator combination variety. 

\subsubsection{Model Execution Time (Efficiency)}
\label{RQ5_efficiency}

\begin{figure*}[htpb]
    \centering
    \includegraphics[width=0.9\textwidth]{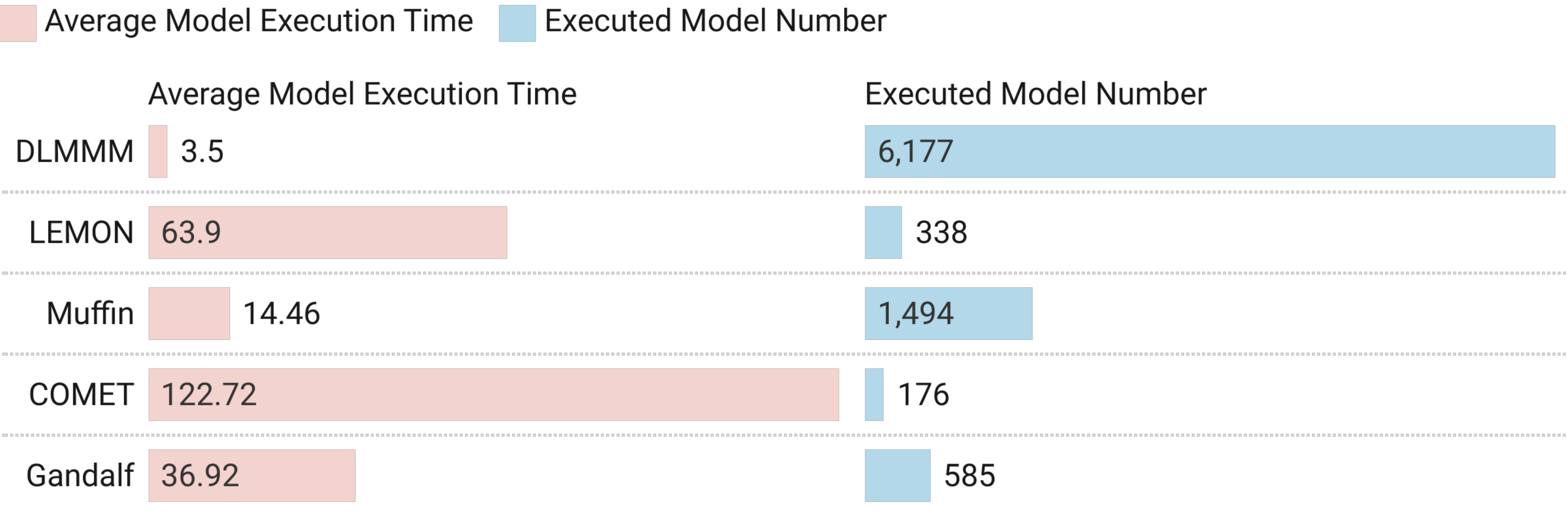}
    \caption{Comparison between DLMMM and Baselines in Efficiency}
    \label{chart_efficiency}
\end{figure*}

To evaluate the efficiency of DLMMM, we count the number of models executed by DLMMM in six hours and calculate the average model execution time. The experimental results are shown in Table \ref{table_efficiency}. In this table, the first column represents the adopted dataset. The second column represents the average model execution time. The last column represents the number of executed test input models. In addition, we calculate the average number of executed test input models on all datasets and compare it with four baselines, as shown in Figure \ref{chart_efficiency}. In this figure, the first column represents the method name, and the meaning of the following two columns is the same as in Table \ref{table_efficiency}. The experimental results show that, under all datasets, DLMMM takes 5.06 seconds at most and 3.50 seconds on average to execute each single model. However, LEMON, Muffin, COMET, and Gandalf take 63.90 seconds, 14.46 seconds, 122.72 seconds, and 36.92 seconds, respectively. In conclusion, the time consumption of DLMMM to execute a single model is within 35\% of all state-of-the-art methods.
\begin{table*}[htpb]
  \caption{Efficiency of DLMMM}
  \label{table_efficiency}
  \centering
  \begin{tabular}{ccc}
  \hline
   & \textbf{Average Model Execution Time} & \textbf{Executed Model Number}\\
  \hline
    Random & 4.51 & 4,785 \\
    MNIST & 3.06 & 7,051 \\
    Fashion-MNIST & 3.30 & 6,550 \\
    CIFAR-10 & 5.06 & 4,268 \\
    ImageNet & 2.74 & 7,876 \\
    Sine-Wave & 2.76 & 7,813 \\
    Stock-Price & 4.41 & 4,900 \\
    \hline
  \end{tabular}
\end{table*}

\begin{center}
\fcolorbox{black}{lightgray}{\parbox{.95\linewidth}{Answer to RQ5: DLMMM successfully generates test input models with higher operator combination variety and less model execution time. Compared to the best performance of all baselines, DLMMM achieves 250.13\% improvement in operator combination variety. The average model execution time is within 35\% of all state-of-the-art methods.}}
\end{center}

\subsection{RQ6: Ablation Study on Heuristic Guidance}
\label{RQ6_Ablation_Study}
Heuristic guidance is the core component of DLMMM. In this section, we conduct an ablation study to evaluate the contribution of our heuristic guidance, measurement fusion, and each model measurement. Specifically, we construct four baselines. Firstly, we design a baseline named $DLMMM_{r}$. $DLMMM_{r}$ disables all heuristic guidance in DLMMM and randomly generates test input models. Secondly, we design a baseline for each model measurement, respectively named $DLMMM_{p}$ (designed for bug detection effectiveness), $DLMMM_{o}$ (designed for operator combination variety), and $DLMMM_{t}$ (designed for model execution time). In these baselines, the measurement fusion is disabled, and only the corresponding model measurement is included into the heuristic guidance.

We respectively run DLMMM and all baselines for six hours and record detected crashes and NaN \& inconsistency bugs. For fairness, all test input tensors are randomly generated and kept the same among DLMMM and all baselines. The experimental results are shown in Table \ref{table_ablation_study}. In this table, the first column represents the method name. The following two columns represent the number of detected crashes and NaN \& inconsistency bugs. From the experimental results, we have the following conclusions. Firstly, DLMMM (detecting 19 crashes and 175 NaN \& inconsistency bugs) detects more bugs than $DLMMM_r$ (detecting no crash and one NaN \& inconsistency bug), which shows the positive contribution of our heuristic guidance. Secondly, DLMMM detects more bugs than $DLMMM_p$ (detecting one crash and eight NaN \& inconsistency bugs), $DLMMM_o$ (detecting five crashes and two NaN \& inconsistency bugs), and $DLMMM_t$ (detecting one crash and three NaN \& inconsistency bugs), which shows the positive contribution of our measurement fusion. Thirdly, $DLMMM_p$, $DLMMM_o$ and $DLMMM_t$ all detect more bugs than $DLMMM_r$, which respectively shows the positive contribution of bug detection effectiveness, operator combination variety, and model execution time.

\begin{table}[htpb]
  \caption{Comparison in Effectiveness in Ablation Study} 
  \label{table_ablation_study}
  \centering
  \begin{tabular}{ccc}
  \hline
  &\textbf{Crash} & \textbf{NaN \& Inconsistency}\\  \hline
    $DLMMM$ & 19 & 175\\
    $DLMMM_r$ & 0 & 1\\
    $DLMMM_p$ & 1 & 8\\
    $DLMMM_o$ & 5 & 2\\
    $DLMMM_t$ & 1 & 3\\
    \hline
  \end{tabular}
\end{table}

To investigate whether DLMMM's designs in heuristic guidance contribute to its testing effectiveness beyond simply generating a larger number of test input models. We record the number of bugs detected by DLMMM when it generates the same number of test input models as existing methods. The experimental results are shown in Table \ref{table_ablation_study_2}. The first column of the table indicates the number of generated test input models. The first four rows show the results when DLMMM generats the same number of models as COMET (176), LEMON (338), Gandalf (585), and Muffin (1494), respectively. The fifth row shows the results using the total number of models generated by DLMMM. The following two columns indicate the number of crashes and NaN \& inconsistency bugs detected by DLMMM.
As shown in the table, when generating 176 models, DLMMM detects 15 crashes and 22 NaN \& inconsistency bugs. In contrast, as shown in Figure \ref{chart_crash_detection} and \ref{chart_inconsistency_detection}, COMET detects only 13 crashes and 6 NaN \& inconsistency bugs when generating the same number of models (176). When generating 338 models, DLMMM detects 17 crashes and 30 NaN \& inconsistency bugs. However, LEMON detects only 1 crash and 0 NaN \& inconsistency bugs with the same model count (338), as shown in Figure \ref{chart_crash_detection} and \ref{chart_inconsistency_detection}. When generating 585 models, DLMMM detects 18 crashes and 43 NaN \& inconsistency bugs, whereas Gandalf detects only 11 crashes and 36 NaN \& inconsistency bugs under the same condition. When generating 1494 models, DLMMM detects 18 crashes and 66 NaN \& inconsistency bugs. However, Muffin does not detect any bugs when generating the same number of models (1494), as shown in Figure \ref{chart_crash_detection} and \ref{chart_inconsistency_detection}.
These experimental results demonstrate that DLMMM detects more crashes and NaN \& inconsistency bugs than the existing methods when generating the same number of test input models. In summary, besides generating more test input models, the heuristic guidance in DLMMM also contributes to improving testing effectiveness under an equal number of test input models.

\begin{table}[htpb]
  \caption{DLMMM's Testing Effectiveness When Generating Different Number of Test Input Models} 
  \label{table_ablation_study_2}
  \centering
  \begin{tabular}{ccc}
  \hline
\textbf{Generated Model Number}&\textbf{Crash} & \textbf{NaN \& Inconsistency}\\  \hline
    176 & 15 & 22\\
    338 & 17 & 30\\
    585 & 18 & 43\\
    1494 & 18 & 66\\
    6177 & 19 & 175\\
    \hline
  \end{tabular}
\end{table}

\begin{center}
\fcolorbox{black}{lightgray}{\parbox{.95\linewidth}{Answer to RQ6: The ablation study shows that DLMMM's heuristic guidance, operator fusion, and each model measurement all contribute to improving DLMMM's bug detection effectiveness. Besides generating more test input models, DLMMM's above designs in heuristic guidance also contributes to improving testing effectiveness under an equal number of the test input models.}}
\end{center}
\section{Validity \& Threat}

In this section, we discuss the threats of DLMMM. Overall, there are three threats that may affect the validity of our method: the internal threat, the external threat, and the construct threat.

The internal threat lies in the model execution time, which may fluctuate on different devices. To eliminate this threat, we execute all methods on the same device.

The external threat lies in the DL frameworks and operators tested. To eliminate this threat of DL framework selection, we apply our method to test three widely used frameworks, respectively TensorFlow, PyTorch and MindSpore. The other frameworks can also be tested by applying the core idea of our method. Additionally, with the development of large language models (LLMs), developers have encapsulated some commonly used model substructures as operators (e.g., the substructure called Transformer) to facilitate more convenient model construction. To adapt to this trend and enhance the scalability of DLMMM, the measurement of mutation and operator variety in DLMMM is also designed to be applicable to commonly used model substructures (e.g., LSTM and Transformer are also tested operators in DLMMM), rather than simply limited to individual functional operators (such as the operator $Convolution$).

The construct threat lies in the randomness in our experiment. The randomness in differential testing may affect the accuracy of our experiment. To reduce the impact of randomness, we generate a larger number of models than all baselines. For example, we run our method for six hours, generating more than 4000 models on each dataset, which are the most in existing methods (338 in LEMON, 1494 in Muffin, 176 in COMET, and 585 in Gandalf). 
\section{Related Works}
DL framework bugs have received extensive attention in recent years. Many empirical studies on DL framework bugs are conducted, which establish the foundation for DL framework testing \cite{DLbug1,DLbug2,DLbug3,DLbug4,DLbug5,DLbug6,DLbug7,DLbug8,DLbug9}.
DL framework testing basically contains interface-based DL framework testing and model-based DL framework testing. Interface-based DL framework testing separately invokes each operator provided by DL frameworks. Among them, Predoo \cite{predoo} mutates test input tensors to maximize output precision error. FreeFuzz \cite{freefuzz} conducts
automated API-level fuzzing via mining from actual code and model executions in the wild. DocTer \cite{DocTer} analyzes API documentation to extract input constraints for each API function provided by DL frameworks. DeepREL \cite{DeepREL} infers potential API relations based on their syntactic or semantic information, and synthesizes test programs for invoking relational APIs. Duo \cite{duo} generates test input tensors by employing nine mutation rules derived from genetic algorithms. EAGLE \cite{eagle} invokes equivalent operators to test DL frameworks. The above methods successfully trigger bugs in a single operator, but ignore some bugs which can only be triggered by invoking operator combinations in DL models. To better detect DL framework bugs, some researchers propose methods which regard DL models as test inputs. Among them, Cradle \cite{cradle} collects publicly available models and directly uses them to test DL frameworks. However, the number of publicly available models is limited. To overcome this challenge, researchers propose various methods to generate test input models. These methods are mainly categorized into generation-based and mutation-based methods. Generation-based methods construct DL models from scratch. Among them, Gandalf \cite{gandalf} generates new models using context-free grammar and optimizes model generation using deep-Q network. Audee \cite{audee} randomly generates a set of DL models by combining different parameter configurations in DL frameworks. Muffin \cite{muffin} first designs multiple templates, which determines the connections between operators. Then, Muffin generates new models by determining the types of operators in each position of the templates. DLLEN \cite{dllen}, Fuzzgpt \cite{fuzzgpt}, and TitanFuzz \cite{titanfuzz} generate DL models with the aid of large language models. Mutation-based methods generate new DL models by mutating seed models. Specifically, LEMON \cite{lemon} chooses publicly available models as seeds, and designs multiple mutation rules to modify the model structure, such as adding operators, deleting operators, and connecting operators. ModelMeta \cite{ModelMeta} designs Metamorphic Relations (MRs) based on input data and parameter settings to generate equivalent test input models. DevMuT \cite{mu2024devmut} designs model mutation rules based on the developer expertise. DLJSFuzzer \cite{DLJSFuzzer} designs mutation rules to generate JavaScript test input models. COMET \cite{comet} adopts the MH algorithm, a Markov Chain Monte Carlo (MCMC) algorithm to select mutation rules. In this work, we apply a mutation-based method to generate DL models.

The role of heuristic guidance differs in different methods. Specifically, Muffin \cite{muffin} adopts heuristic guidance to control both the selection of templates and the selection of specific operators for each position within templates. LEMON \cite{lemon} and COMET \cite{comet} adopt heuristic guidance to control the selection of mutation seeds and mutation rules. However, existing methods fail to guide the specific implementation of each mutation rule (e.g., controlling the probability of inserting each type of operator). To overcome this challenge, we design multiple-level heuristic guidance for test input model generation, which can not only heuristically control the probability of each mutation rule being selected but also control the specific implementation of the mutation rules.

\section{Conclusion}

In this paper, we propose DLMMM, a DL framework testing method via heuristic guidance based on multiple model measurements. DLMMM is the first to include multiple model measurements into the heuristic guidance in DL framework testing. DLMMM designs and adopts quantitative measurements for bug detection effectiveness, operator combination variety, and model execution time. Based on these measurements, DLMMM designs a measurement fusion to achieve the overall optimization in DL framework testing. To further improve the effectiveness of heuristic guidance, DLMMM designs a multi-level heuristic test input model generation. We evaluate the performance of DLMMM by testing three widely used frameworks and the experimental results show that DLMMM can detect crashes, NaN bugs, and inconsistency bugs effectively. In future work, we aim to further optimize the test input model generation for DL framework testing.



\section*{Acknowledgement}
This work is partially supported by the National Natural Science Foundation of China (U24A20337, 62372228), the Shenzhen-Hong Kong-Macau Technology Research Programme (Type C) (Grant No.SGDX20230821091559018), 
the Open Project of State Key
Laboratory for Novel Software Technology at Nanjing University (Grant No. KFKT2024B21) 
and the Fundamental Research Funds for the Central Universities (14380029).

\bibliographystyle{acm}
\bibliography{sample-base}

\end{document}